\documentclass[twocolumn, pra, showpacs,superscriptaddress,floatfix]{revtex4}
\usepackage{graphicx}
\usepackage{dcolumn}
\usepackage{bm}
\usepackage{amsmath}

\begin{document}

\title{Effective super Tonks-Girardeau gases as ground states of strongly
attractive multi-component fermions}
\author{Xiangguo Yin}
\affiliation{Institute of Physics, Chinese Academy of Sciences, Beijing 100190, China}
\author{Xi-Wen Guan}
\affiliation{Department of Theoretical Physics, Research School of Physics and
Engineering, Australian National University, Canberra ACT 0200, Australia}
\author{M T Batchelor}
\affiliation{Department of Theoretical Physics, Research School of Physics and
Engineering, Australian National University, Canberra ACT 0200, Australia}
\affiliation{Mathematical Sciences Institute, Australian National University, Canberra
ACT 0200, Australia}
\author{Shu Chen}
\affiliation{Institute of Physics, Chinese Academy of Sciences, Beijing 100190, China}

\begin{abstract}
In the strong interaction limit, attractive fermions with $N$-component
hyperfine states in a one-dimensional waveguide form unbreakable bound
cluster states. We demonstrate that the ground state of strongly attractive
SU($N$) Fermi gases can be effectively described by a super Tonks-Girardeau
gas-like state composed of bosonic cluster states with strongly attractive
cluster-cluster interaction for even $N$, and a Fermi duality of a super
Tonks-Girardeau gas-like state composed of fermionic cluster states with
weakly interacting cluster-cluster p-wave interaction for odd $N$.
\end{abstract}

\pacs{03.75.Ss, 05.30.Fk}
\date{\today }
\maketitle

%\email{schen@aphy.iphy.ac.cn}

\section{Introduction}

The experimental progress on trapping ultracold atoms in tightly confined
waveguides in a well-controlled way \cite{gorlitz,esslinger,Paredes,Toshiya} 
has stimulated intensive study of the physical properties of
one-dimensional (1D) quantum gases. The effective interaction strength
between atoms in a 1D waveguide can be tuned via Feshbach resonance or
confinement-induced resonance \cite{Olshanii}, leading to the experimental
realization of Tonks-Girardeau (TG) gases \cite{Paredes,Toshiya}. The TG gas
describes the strongly repulsive Bose gas \cite{Girardeau,Lieb}. Starting
from the TG gas and then switching the interaction between atoms from
strongly repulsive to strongly attractive, the experimental realization of a
1D \emph{super} Tonks-Girardeau (STG) gas of bosonic Cesium atoms was
reported very recently \cite{Haller}. In contrast to the TG gas, the STG gas
describes a gas-like phase of the attractive Bose gas \cite{Astrakharchik1,
Astrakharchik2,Batchelor}, which is metastable against falling into its
cluster-type ground state \cite{McGuire}, despite the fact that the
interaction between atoms is strongly attractive \cite{Chen1,Girardeau_STG}.

Although the STG gas realized in \cite{Haller} is a metastable highly
excited state of the attractive Bose gas, in recent theoretical work \cite%
{Chen2} it was found that the ground state of a strongly attractive spin-1/2
Fermi gas can be effectively described by the STG gas. Intuitively, two
fermions with opposite spins form a tightly bound state and the bound pairs
of fermions can be viewed as composite bosons with a mass of $2m_{F}$. The
effective interaction between the composite bosons is also attractive with
the interaction strength given by $c_{B}=2c_{F}$ \cite{Chen2,Egger05}.
Conversely, the ground state of the bound Fermi pairs is described by the
STG phase of attractive bosons \cite{Chen2}. Very recently, 
multi-component Fermi gases have attracted considerable 
interest \cite{Hu,Guan10,SUN} due to the novel existence of different 
sizes of molecules. Here we consider the interesting question of whether 
the ground state of a strongly attractive multi-component fermionic system 
can be also effectively described by a STG gas of multi-particle bound states. 
A positive answer is confirmed by explicit identification of the general mapping 
relation between the attractive SU($N$) Fermi gas and the STG gas.

\section{attractive $N$-bound fermions}

We consider a delta-function interacting system composed of $N_{F}$ atomic
attractive fermions with equal mass $m_{F}$ which occupy $N$ hyperfine
levels with identical particle numbers $N^{i}=N_{N}=N_{F}/N$ $\left(
i=1,\ldots ,N\right) $ and constrained by periodic boundary conditions to a
line of length $L$. If the interactions are spin independent, the
Hamiltonian reads
\begin{equation}
H_{F}=\sum_{i=1}^{N_{F}}-\frac{\hbar ^{2}}{2m_{F}}
\frac{\partial ^{2}}{\partial x_{i}^{2}}+g_{F}\sum_{i<j}\delta (x_{i}-x_{j}),  \label{HF}
\end{equation}
where $g_{F}=-2\hbar ^{2}/(m_{F}{a_{\mathrm{1D}}^{F}})$ is the interaction
strength with ${a_{\mathrm{1D}}^{F}}$ the 1D effective s-wave scattering length. 
Although the interactions in Hamiltonian (\ref{HF}) are 
represented in terms of delta interaction, the exchange antisymmetric
wavefunction for fermions gives the restriction that interactions among the
same fermion level are prohibited. Different symmetries of the wavefunction
produce different Bethe ansatz equations even if the Hamiltonian has the
same form as Eq.~(\ref{HF}) \cite{Sutherland,YangK}. For simplicity,
we use the dimensionless coupling constant $\gamma _{F}=c_{F}/n_{F}$ with
density $n_{F}={N}_{F}/{L}$ and $c_{F}=-2/{a_{\mathrm{1D}}^{F}}$. In the
strongly attractive limit $\left( g_{F}\rightarrow -\infty \right) $, atoms
form tightly bound states with each bound state composed of $N$ fermions in
different hyperfine states \cite{Takahashi}. No tightly bound state with more 
than $N$ fermions can be formed in an $N$-component Fermi gas due to 
the Pauli exclusion principle.

If $N$ is even, the tightly bound state can be viewed as a composite boson
with a mass $m_{B}=Nm_{F}$. In this work, we find that the ground state of
the strongly attractive SU($N$) Fermi gas with $N$ {even} can be effectively
described by a STG gas of attractive composite bosons. This can be viewed as
a direct generalization of the SU(2) Fermi gas result \cite{Chen2}. On the
other hand, if $N$ is odd, the tightly bound state can only be viewed as a
composite fermion. Obviously, strongly attractive SU($N$) fermions with $N$
odd cannot be mapped to an effective bosonic gas as for the two-component
case \cite{Chen2}. Nevertheless, we shall show that it can be mapped to a
spinless Fermi gas with weakly interacting p-wave interaction. Due to a
general Fermi-Bose mapping \cite{Cheon}, eigenstates of a spinless Fermi gas
with p-wave interaction of any strength can be mapped to those of a 1D Bose
gas with delta-function interactions. Therefore the gas-like state of a
weakly repulsive p-wave Fermi gas can be viewed as the Fermi correspondence
of the STG phase of a strongly attractive Bose gas.

Before construction of the mapping between the ground state of the strongly
attractive SU($N$) Fermi gas and the STG phase of composite bosons or
fermions, we first discuss the solution of the SU($N$) Fermi gas which is
exactly solvable by the Bethe ansatz method. The eigenvalues of Hamiltonian (%
\ref{HF}) are given by
\begin{equation}
E=\frac{\hbar ^{2}}{2m_{F}}\sum_{j=1}^{N_{F}}k_{j}^{2}
\end{equation}%
with quasi-momentum $k_{j}$ determined by the Bethe ansatz equations (BAEs)
\cite{Sutherland,YangK,Takahashi}
\begin{eqnarray}
&&\hspace{10mm}\exp (\mathrm{i}k_{j}L)=\prod_{\alpha =1}^{M_{1}}\frac{%
k_{j}-\Lambda _{\alpha }^{(1)}+\mathrm{i}c_{F}^{\prime }}{k_{j}-\Lambda
_{\alpha }^{(1)}-\mathrm{i}c_{F}^{\prime }},  \label{BE1} \\
&&\prod_{\beta =1}^{M_{\ell -1}}\frac{\Lambda _{\alpha }^{(\ell )}-\Lambda
_{\beta }^{(\ell -1)}+\mathrm{i}c_{F}^{\prime }}{\Lambda _{\alpha }^{(\ell
)}-\Lambda _{\beta }^{(\ell -1)}-\mathrm{i}c_{F}^{\prime }}=-\prod_{\gamma
=1}^{M_{\ell }}\frac{\Lambda _{\alpha }^{(\ell )}-\Lambda _{\gamma }^{(\ell
)}+\mathrm{i}c_{F}}{\Lambda _{\alpha }^{(\ell )}-\Lambda _{\gamma }^{(\ell
)}-\mathrm{i}c_{F}}  \notag \\
&&\hspace{10mm}\times \prod_{\nu =1}^{M_{\ell +1}}\frac{\Lambda _{\alpha
}^{(\ell )}-\Lambda _{\nu }^{(\ell +1)}-\mathrm{i}c_{F}^{\prime }}{\Lambda
_{\alpha }^{(\ell )}-\Lambda _{\nu }^{(\ell +1)}+\mathrm{i}c_{F}^{\prime }},
\label{BE2}
\end{eqnarray}%
for $j=1,\ldots ,N_{F}$, $\alpha =1,\ldots ,M_{\ell }$ and $\ell =1,\ldots,N-1$. 
We have denoted $M_{0}=N_{F}$ and $\Lambda _{j}^{(0)}=k_{j}$. The
parameters $\left\{ \Lambda _{\alpha }^{(\ell )}\right\} $ are the spin
rapidities. The quantum numbers are given by $M_{l}=\left( N-l\right) N_{N}$
and $c_{F}^{\prime }=c_{F}/2$.

For strongly attractive attraction, i.e., $L|c_{F}|\gg 1$, the BAEs permit
different sizes of charge bound state. As a consequence of the Pauli 
exclusion principle and the $SU(N)$ symmetry, there is no tightly
bound state with more than $N$ fermions for the $SU(N)$ Fermi gas \cite{Gu-Yang}. 
For the ground state, there are equal numbers of particles in each hyperfine spin state. 
In this state, the charge bound state in $k$-space is of the form 
\begin{equation}
k_{q,j}=\Lambda _{q}^{(N-1)}+\left( N+1-2j\right) c_{F}^{\prime }+
O\left(\mathrm{i}\delta |c_{F}|\right),
\end{equation}%
for $j=1,2,\ldots ,N$ and $q=1,2,\ldots ,N_{F}/N$. The spin rapidities form a 
certain pattern of spin string solutions. For the ground state, each charge
bound state $k_{q,j}$ with different $q$ is accompanied by a 
spin string $\left\{ \Lambda _{q,\alpha }^{(1)}\right\} $ of length $N-1$with 
$\alpha=1,2,\ldots ,N-1$, a spin string $\left\{ \Lambda _{q,\alpha}^{(2)}\right\} $ 
of length $N-2$ with $\alpha =1,2,\ldots ,N-2$, and so on, until the real root
$\Lambda _{q}^{(N-1)}$ in the last spin branch \cite{Takahashi}. 
In this special case, the spin strings read
\begin{equation}
\Lambda _{q,\alpha }^{\left( r\right) }=\Lambda _{q}^{(N-1)}+\mathrm{i}%
\left( N-r+1-2\alpha \right) c_{F}^{\prime }+O\left( \mathrm{i}\delta
|c_{F}|\right) ,
\end{equation}%
with $\alpha =1,\ldots ,N-r$ for $r=1,...,N-2$, respectively. In the above equations 
$\delta$ is a very small number of order $\exp (-L|c_{F}|)$. 

Substituting the charge
bound states $k_{q,j}$ with $j=1,2,\ldots ,N$ and the spin strings into Eq.~(\ref{BE1}) 
results in $N$ equations. After multiplying these $N$ equations
together and combining with Eq.~(\ref{BE2}) (see Appendix A) the BAEs reduce to
\begin{equation}
\exp \left( N\mathrm{i}\Lambda _{q}L\right) =\left( -1\right)
^{N_{F}-1}\prod_{\beta =1}^{N_{N}}\prod_{r=1}^{N-1} \frac{\Lambda
_{q}-\Lambda _{\beta }+\mathrm{i}rc_{F}}{\Lambda _{q}-\Lambda _{\beta }-%
\mathrm{i}rc_{F}}.  \label{BAE2}
\end{equation}%
The eigenvalues of Hamiltonian (\ref{HF}) are then given by
\begin{equation}
E=-N_{N}\epsilon _{b}+\frac{\hbar ^{2}}{2m_{F}}\sum_{q=1}^{N_{N}}N\Lambda
_{q}^{2}
\end{equation}%
with binding energy $\epsilon _{b}=\left( \hbar ^{2}/2m_{F}\right) N\left(
N^{2}-1\right) c_{F}^{2}/12$.

In the strongly attractive limit and in the absence of an external field,
the $N$-fermion clusters are unbreakable and we may subtract the binding
energy from the energy, i.e.,
\begin{equation}
E_{F}=E+N_{N}\epsilon _{b}= \frac{\hbar ^{2}} {2m_{F}} \sum_{q=1}^{N_{N}}N%
\Lambda _{q}^{2},
\end{equation}%
which includes the kinetic energy of the bound clusters and the interaction
energy produced from cluster-cluster scattering.

In the thermodynamic limit, $N_{F}\rightarrow \infty $ and $L\rightarrow
\infty $ at fixed density $n_{F}$, the energy of the system can be
represented in the integral form
\begin{equation}
E_{F}=\frac{\hbar ^{2}L}{2m_{F}}N\int_{-B}^{B}\Lambda ^{2}\rho _{F}\left(
\Lambda \right) d\Lambda  \label{energy1}
\end{equation}%
where $\rho _{F}\left( \Lambda \right) $ is the density distribution for $%
\Lambda $ determined by the integral form of BAEs (\ref{BAE2}) as
\begin{equation}
\rho _{F}\left( \Lambda \right) =\frac{N}{2\pi }-\frac{1}{\pi }%
\sum_{r=1}^{N-1}\int_{-B}^{B}\frac{r\left\vert c_{F}\right\vert }{%
r^{2}c_{F}^{2}+\left( \Lambda -\Lambda ^{\prime }\right) ^{2}}\rho
_{F}\left( \Lambda ^{\prime }\right) d\Lambda ^{\prime }.  \label{IBAE}
\end{equation}%
The integration limit $B$ is determined by the linear density $%
n_{F}=N\int_{-B}^{B}\rho _{F}\left( \Lambda \right) d\Lambda $.

In terms of the dimensionless energy $e_{N}\left( \gamma _{F}\right) $, we
have
\begin{equation}
E_{F}=\frac{\hbar ^{2}L}{2m_{N}}n_{N}^{3}e_{N}\left( \gamma _{F}\right) ,
\end{equation}%
where $n_{N}=n_{F}/N$, $m_{N}=Nm_{F}$ and
\begin{equation}
e_{N}\left( \gamma _{F}\right) =\frac{N^{6}\left\vert \gamma _{F}\right\vert
^{3}}{\lambda ^{3}}\int_{-1}^{1}z^{2}g_{N}\left( z\right) dz.
\end{equation}%
Here we have defined $z=\Lambda /B$, $\lambda =\left\vert c_{F}\right\vert
/B $ and $g_{N}\left( z\right) =\rho \left( Bz\right) /N$. The scaled
density distribution $g_{N}\left( z\right) $ is determined by
\begin{eqnarray}
g_{N}\left( z\right) &=&\frac{1}{2\pi }-\frac{1}{\pi }\sum_{r=1}^{N-1}%
\int_{-1}^{1}\frac{r\lambda }{r^{2}\lambda ^{2}+\left( z-z^{\prime }\right)
^{2}}g_{N}\left( z^{\prime }\right) dz^{\prime },  \notag \\
\lambda &=&N^{2}\left\vert \gamma _{F}\right\vert \int_{-1}^{1}g_{N}\left(
z\right) dz,  \notag  \label{BA_num}
\end{eqnarray}%
which come from Eq.~(\ref{IBAE}) and the linear density $n_{F}$. In the
strongly attractive limit with $|\gamma _{F}|\gg 1$, we can expand the
dimensionless energy in terms of $1/|\gamma _{F}|$. Up to 3rd order, this
gives
\begin{eqnarray}
E_{F} &=&\frac{\hbar ^{2}N_{N}^{3}}{2m_{N}L^{2}}\frac{\pi ^{2}}{3}\left[ 1+%
\frac{4}{\left\vert \gamma _{N}\right\vert }+\frac{12}{\gamma _{N}^{2}}%
\right.  \notag \\
&&\left. +\frac{32}{\left\vert \gamma _{N}\right\vert ^{3}}\left( 1-\frac{%
\pi ^{2}\eta }{15\zeta ^{3}}\right) \right] ,  \label{EF}
\end{eqnarray}%
\newline
in which $\gamma _{N}=N^{2}\gamma _{F}/\zeta $, with $\zeta
=\sum_{r=1}^{N-1}1/r$ and\ $\eta =\sum_{r=1}^{N-1}1/r^{3}$.

\subsection{Equivalence to a super TG gas for $N$ even}

The unbreakable $N$-fermion cluster state is effectively described as a
composite boson for even $N$. Before constructing the mapping relation, we
first give a brief review of the STG state of the attractive Bose gas. The
1D interacting Bose gas composed of $N_{B}$ bosons with mass $m_{B}$ is
described by the Hamiltonian
\begin{equation}
H_{B}=\sum_{i=1}^{N_{B}}- \frac{\hbar ^{2}}{2m_{B}}\frac{\partial ^{2}}{%
\partial x_{i}^{2}}+g_{B}\sum_{i<j}\delta (x_{i}-x_{j}),  \label{Hb}
\end{equation}%
with interaction $g_{B}=-2\hbar ^{2}/(m_{B}a_{\mathrm{1D}}^{B})$ where $a_{%
\mathrm{1D}}^{B}$ is the 1D s-wave scattering length. The energy eigenvalues
are given by
\begin{equation}
E_{B}=\frac{\hbar ^{2}}{2m_{B}}\sum_{j=1}^{N_{B}}k_{j}^{2},
\end{equation}
where the $k_{j}$ are determined by the BAE \cite{Lieb}
\begin{equation}
\exp \left( \mathrm{i}k_{j}L\right) =-\prod_{l=1}^{N_{B}}\frac{k_{j}-k_{l}+%
\mathrm{i}c_{B}}{k_{j}-k_{l}-\mathrm{i}c_{B}},  \label{BAEbose}
\end{equation}%
with $c_{B}=m_{B}g_{B}/\hbar ^{2}=-2/a_{\mathrm{1D}}^{B}$.

For attractive bosons, the ground state solution for the BAE (\ref{BAEbose})
is a complex string solution corresponding to McGuire's cluster state \cite%
{McGuire}. On the other hand, the BAE (\ref{BAEbose}) has real solutions
even for $c_B<0$, which correspond to some highly excited states of the
attractive Bose gas. The super TG state is the lowest gas-like state with
real solutions for BAE (\ref{BAEbose}) \cite{Batchelor,Chen1,Chen2}. In the
strongly attractive limit, the energy of the STG state of the attractive
Bose gas can be expressed as
\begin{eqnarray}
E_{STG}&=& \frac{\hbar ^{2}N_{B}^{3}}{2m_{B}L^{2}}\frac{\pi ^{2}}{3}\left[ 1+%
\frac{4}{\left\vert \gamma _{B}\right\vert }+\frac{12}{\left\vert \gamma
_{B}\right\vert ^{2}} \right.  \notag \\
& & \left. +\frac{32}{\left\vert \gamma _{B}\right\vert ^{3}}\left( 1-\frac{%
\pi ^{2}}{15}\right) \right]  \label{ESTG}
\end{eqnarray}%
with $\gamma _{B}=c_{B}/n_{B}$.

Comparing equations (\ref{EF}) and (\ref{ESTG}), it is clear that the two
expressions are identical up to the second order of $\gamma _{F}$ if $\gamma
_{B}=\gamma_{N}=N^{2}\gamma _{F}/\zeta $, $N_{B}=N_{N}=N_{F}/N$ and $%
m_{B}=m_{N}=Nm_{F}$. Since the $N$-bound state formed by $N$ fermions with
opposite spin for even $N$ has a mass $m_{B}=Nm_{F}$, we can conclude that
the $N_{N}$ $N$-bound states are equivalently described by the super-TG
phase of the interacting Bose gas with the effective 1D scattering length
\begin{equation}
a_{\mathrm{1D}}^{B}=\frac{\zeta }{N}{a_{\mathrm{1D}}^{F}}.  \label{aB}
\end{equation}

Schematically, we illustrate such a mapping in Fig. 1(a) by taking the SU(4)
Fermi gas as an example. We also compare the ground state energy of the SU($%
N $) ($N=2,4$) Fermi gas with the energy of the STG phase of the Bose gas
composed of composite bosons with mass $Nm_F$ in Fig. \ref{fig2}. The ground
state energy of the SU(2) Fermi gas is identical to the energy of the STG
phase of the Bose gas for all $\gamma_{N}$ \cite{Chen2}, whereas the ground
state energy of the SU(4) Fermi gas matches very well to that of the STG gas
for large $\gamma _{N}$. As shown in Fig. 3, the relative error for the
ground state energy of the SU($N$) Fermi gas and the energy of the
corresponding STG gas is less than 0.1 for $|\gamma _{N}|=10$, and less than
$10^{-7}$ for $|\gamma _{N}|=600$. This indicates that the STG gas provides
a good effective description for the ground state of the strongly attractive
SU($N$) Fermi gas, although the mapping is not exact for all $\gamma_{N}$
like for the SU(2) case.

\begin{figure}[tbp]
\includegraphics[width=1.0\linewidth]{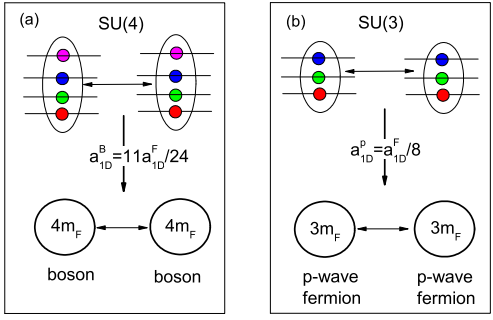}
\caption{(Color online) The strongly attractive $N$-bound state Fermi gas
can be effectively described by a super Tonks-Girardeau gas composed of
attractive bosons for (a) even $N$ and can be effectively described by a
super Fermi Tonks-Girardeau gas composed of p-wave repulsive fermions for
(b) odd $N$. }
\label{fig1}
\end{figure}

\begin{figure}[tbp]
\includegraphics[width=1.0\linewidth]{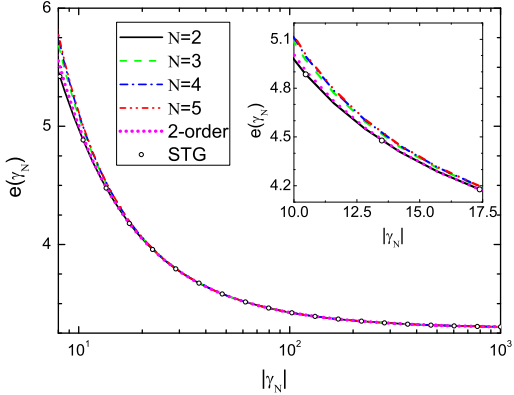}
\caption{(Color online) Comparison of the ground state energies of the
attractive SU($N$) Fermi gas and the energy of the corresponding STG phase
of bosons (for even $N$) or p-wave fermions (for odd $N$) (effective
repulsive p-wave fermions) for different $N$ with large interaction $\protect%
\gamma _{N}$. The inset shows a magnified view of ground state energies.}
\label{fig2}
\end{figure}

\begin{figure}[tbp]
\includegraphics[width=1.0\linewidth]{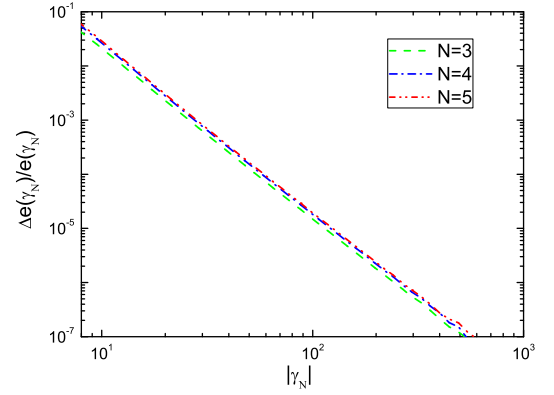}
\caption{(Color online) The relative error vs $|\protect\gamma_N|$, where $%
\Delta e\left( \protect\gamma_{N}\right) =e\left( \protect\gamma _{N}\right)
-e_{STG}\left( \protect\gamma_{N} \right) $.}
\label{fig3}
\end{figure}

\subsection{Equivalence to a super Fermi TG\ gas for $N$ odd}

The Hamiltonian for the 1D p-wave interacting polarized Fermi gas reads
\begin{equation}
H_{p}=\sum_{i=1}^{N_{p}}-\frac{\hbar ^{2}}{2m_{p}}\frac{\partial ^{2}}{%
\partial x_{i}^{2}}+g_{p}\sum_{i<j}V(x_{i}-x_{j}),  \label{Hp}
\end{equation}%
where $V(x_{i}-x_{j})=\left( \frac{\partial }{\partial x_{i}}-\frac{\partial
}{\partial x_{j}}\right) \delta (x_{i}-x_{j})\left( \frac{\partial }{%
\partial x_{i}}-\frac{\partial }{\partial x_{j}}\right) $ is the
pseudo-potential for $p$-wave interaction \cite{Blume,pwave,pwave_Girardeau}
and $g_{p}=-2\hbar ^{2}a_{\mathrm{1D}}^{p}/m_{p}$ \cite{Blume,pwave}. The
dimensionless interaction parameter is defined by $\gamma
_{p}=m_{p}g_{p}n_{p}/\hbar ^{2}=-2a_{\mathrm{1D}}^{p}n_{p}$. For p-wave
interacting fermions, the energy eigenvalues are given by
\begin{equation}
E_{p}=\frac{\hbar ^{2}}{2m_{p}}\sum_{j=1}^{N_{p}}k_{j}^{2},
\end{equation}%
where the quasi-momenta $k_{j}$ are determined by the BAE \cite{Hao,Grosse}
\begin{equation}
\exp \left( \mathrm{i}k_{j}L\right) =-\prod_{l=1}^{N_{p}}\frac{k_{j}-k_{l}+%
\mathrm{i}c_{p}}{k_{j}-k_{l}-\mathrm{i}c_{p}},  \label{BAEpwave}
\end{equation}%
where the parameter $c_{p}=-1/(2a_{\mathrm{1D}}^{p})$.

It is clear that BAE (\ref{BAEpwave}) is identical to BAE (\ref{BAEbose}) if
$c_{p}= c_{B}$ and $N_{p}=N_{B}$ \cite{Hao,Grosse}, which means that there
is a one-to-one correspondence between the p-wave Fermi gas and the
interacting Bose gas \cite{Cheon}. Correspondingly, the STG state of the
attractive Bose gas has a Fermi correspondence which is the lowest gas-like
state of the weakly interacting p-wave fermions with $g_{p} \rightarrow
0^{+} $. For weakly interacting p-wave fermions in the thermodynamic limit,
the energy of the lowest gas-like state has the form
\begin{eqnarray}
E_{p} &=&\frac{\hbar ^{2}}{2m_{p}}\frac{N_{p}^{3}}{L^{2}}\frac{\pi ^{2}}{3}%
\left[ 1+4\left\vert \gamma _{p}\right\vert +12\left\vert \gamma
_{p}\right\vert ^{2}\right.  \notag \\
&&\left. +32\left( 1-\frac{\pi ^{2}}{15}\right) \left\vert \gamma
_{p}\right\vert ^{3}+\cdots \right]  \label{Ep}
\end{eqnarray}%
where $\left\vert \gamma _{p}\right\vert \ll 1$.

Comparing equations (\ref{EF}) and (\ref{Ep}), it is clear that the two
expressions are identical up to the second order in $\gamma _{F}$ if $%
\gamma_{p}=1/\gamma _{N}=\zeta /\left( N^{2}\gamma _{F}\right) $, $%
N_{p}=N_{N}=N_{F}/N$ and $m_{p}=m_{N}=Nm_{F}$. Since the $N$-bound state
formed by $N$ fermions with different hyperfine states for odd $N$ is a
composite fermion with a mass $m_{p}=Nm_{F}$, we can conclude that the $%
N_{N} $ $N$-bound states are equivalently described by the gas-like state of
the weakly interacting p-wave Fermi gas with the effective 1D scattering
length
\begin{equation}
a_{\mathrm{1D}}^{p}=\frac{\zeta a_{\mathrm{1D}}^{F}}{4N}.  \label{ap}
\end{equation}

The mapping is schematically displayed in Fig. 1(b). The comparison of the
energies of the SU($N$) ($N=3,5$) Fermi gas and the STG phase of the p-wave
Fermi gas is also given in Fig. \ref{fig2}, which indicates a good matching
in the limit of large $|\gamma_{N}|$. Similarly, as shown in Fig. 3, the
relative error for the ground state energy of the SU($N$) Fermi gas and the
energy of the corresponding STG state of the p-wave Fermi gas is less than $%
10^{-7}$ for $|\gamma _{N}|=600$. This indicates that, although not exact
for all $\gamma_{N}$, the mapping provides a very good description for large
$|\gamma _{N}|$.

\section{summary}

In summary, we have examined the equivalence between the ground state of the
strongly attractive $N$-component Fermi gas and the super Tonks-Girardeau
phase of an effective Bose or p-wave Fermi gas. By comparing the ground
state energy of strongly attractive fermions with the energy of the super
Tonks-Girardeau phase of the Bose gas or p-wave Fermi gas, we find that the
bound $N$-fermion clusters formed in the strongly attractive regime should
be described by the super Tonks-Girardeau phase of attractive composite
bosons (for even $N$) or composite fermions with effective p-wave
interactions (for odd $N$). The super Tonks-Girardeau gas phase thus
provides an effective description for the ground state of strongly
attractive multi-component fermions.

\textit{Acknowledgments.---} This work was supported by the NSF of China
under Grants No. 10821403 and No. 10974234, programs of CAS, 973 grant No.
2010CB922904 and National Program for Basic Research of MOST. The work of
X.-W.G and M.T.B. has been partially supported by the Australian Research
Council.

\appendix

\section{Derivation of Bethe ansatz equations for Fermi bound states}

Here we show how to derive BAEs for Fermi bound states by taking as example
the system of 3-component fermions. For simplicity, we define the function
\begin{equation*}
e_{n}\left( x\right) =\frac{x+\mathrm{i}nc_{F}^{\prime }}{x-\mathrm{i}%
nc_{F}^{\prime }}.
\end{equation*}
For 3-component fermions  the BAEs (\ref{BE1}) and (\ref{BE2}) are
\begin{eqnarray}
e^{\mathrm{i}k_{j}L}&=&\prod_{\alpha =1}^{M_{1}}e_{1}
\left( k_{j}-\Lambda_{\alpha }^{(1)}\right), \label{A1} \\
\prod_{j=1}^{N_{F}}e_{1}\left( \Lambda _{\alpha }^{(1)}-k_{j}\right) 
&=& -\prod_{\beta =1}^{M_{1}}e_{2}\left( \Lambda _{\alpha }^{(1)}-\Lambda
_{\beta }^{(1)}\right)    \notag \\
&\phantom{=}& \times \prod_{l=1}^{M_{2}}e_{-1}\left( \Lambda _{\alpha }^{(1)}-\Lambda
_{l}^{(2)}\right),   \label{A2} \\
\prod_{\alpha =1}^{M_{1}}e_{1}\left( \Lambda _{l}^{(2)}-\Lambda _{\alpha
}^{(1)}\right) &=&-\prod_{m=1}^{M_{2}}e_{2}\left( \Lambda _{l}^{(2)}-\Lambda
_{m}^{(2)}\right)  \label{A3}
\end{eqnarray}
for $j=1,2,\ldots,N_{F}$, $\alpha ,\beta =1,2,\ldots,M_{1}$ and 
$l,m=1,2,\ldots,M_{2}$. Here we confine our attention to the equally populated case 
 $N_{F}=3N_{3}$, $M_{1}=2N_{3}$ and $M_{2}=N_{3}$.

For strong attraction, i.e., for $L|c_{F}|\gg 1$, the charge bound states and spin strings 
are of the form
\begin{eqnarray}
k_{q,h_{1}}& =&\Lambda _{q}+\mathrm{i}\left( 4-2h_{1}\right) c_{F}^{\prime
}+O\left( \mathrm{i}\delta |c_{F}|\right) , \label{k-b} \\
\Lambda _{q,h_{2}}^{\left( 1\right) }& =& \Lambda _{q}+\mathrm{i}\left(
3-2h_{2}\right) c_{F}^{\prime }+O\left( \mathrm{i}\delta' |c_{F}|\right) \label{L-b},
\end{eqnarray}
for  exponentially small $\delta$ and $\delta'$, 
with $\Lambda _{q}=\Lambda _{q}^{\left( 2\right) }$, $h_{1}=1,2,3$, 
$h_{2}=1,2$, and $q=1,2,\ldots,N_{3}$.

For the attractive regime, the common real parts in the bound states (\ref{k-b}) and spin strings (\ref{L-b}) 
lead to zero factors in the BAEs (\ref{A1}-\ref{A3}). In order to avoid ill-defined equations, we eliminate 
such zero factors in the BAE level by level.  The first step is to deal with the charge bound state $k_{q,h_1}$ 
with $h_1=1,2,3$ in the BAE (\ref{A1}), i.e.
\begin{eqnarray}
 e^{\mathrm{i}3\Lambda_q L}&=& e^{\mathrm{i}(k_{q,1}+k_{q,2}+k_{q,3})L}\nonumber\\
&=&\prod_{h_2=1}^2\prod_{h_1=1}^3e_1\left(k_{q,h_1}-\Lambda^{(1)}_{q,h_2}\right)\nonumber\\
&& \times \prod_{\alpha=1}^{{M_1}/{2}}e_2\left(\Lambda_q-\Lambda_{\alpha}\right) 
e_4\left(\Lambda_q-\Lambda_{\alpha} \right)\label{level1}
\end{eqnarray}
The first terms on the r.h.s. of (\ref{level1}) contain zero factors which have to be eliminated.  
From the second BAE (\ref{A2}), we have 
\begin{eqnarray}
&&\prod_{h_2=1}^2\prod_{h_1=1}^3e_{-1}\left(k_{q,h_1}-\Lambda^{(1)}_{q,h_2}\right)\nonumber\\
&& \times \prod_{\alpha=1}^{{N_F}/{3}}e_2\left(\Lambda_q-\Lambda_{\alpha} \right)e_4\left(\Lambda_q-\Lambda_{\alpha}\right)\nonumber\\
&&= \prod_{h_2=1}^2e_{-1}\left(\Lambda_{q,h_2}^{(1)}-\Lambda_{q}\right)\nonumber\\
 &&\times \prod_{\alpha=1}^{{M_1}/{2}}e_2\left(\Lambda_q-\Lambda_{\alpha}\right)e_4\left(\Lambda_q-\Lambda_{\alpha}\right). 
 \label{level2}
\end{eqnarray}
In order to eliminate the first factor on the r.h.s. of these equations,  we extract this factor from the third BAE (\ref{A3}), i.e.
\begin{eqnarray}
&& \prod_{h_2=1}^2e_{-1}\left(\Lambda_{q,h_2}^{(1)}-\Lambda_{q}\right) 
\prod_{\alpha=1}^{{M_1}/{2}}e_2\left(\Lambda_q-\Lambda_{\alpha} \right)\nonumber\\
 &&=\prod_{m=1}^{M_{2}}e_{2}\left( \Lambda _{l}-\Lambda_{m}\right). \label{level3}
 \end{eqnarray}
Substituting  (\ref{level3}) into  (\ref{level2}), Eq.~(\ref{level1}) becomes 
\begin{eqnarray}
 e^{\mathrm{i}3\Lambda_q L}=\prod_{\alpha=1}^{{M_1}/{2}}e_2\left(\Lambda_q-\Lambda_{\alpha}\right) 
 e_4\left(\Lambda_q -\Lambda_{\alpha} \right), \label{level4}
\end{eqnarray}
which is the 3-component fermion form of Eq.~(\ref{BAE2}).

\end{document}